\documentclass[reprint,aps,prx,floatfix,nofootinbib]{revtex4-2}
\usepackage{newtxtext, newtxmath}  
\usepackage{graphicx, braket, siunitx, mathtools, dcolumn, hyperref}
\usepackage{dsfont}
\usepackage{amsmath}
\usepackage{soul}

\DeclareSIUnit[number-unit-product=]\percent{\char`\%}

\begin{document}

\title{Measurement-based ground state cooling of a trapped ion oscillator}

\author{Chungsun Lee}
\affiliation{Blackett Laboratory, Imperial College London, London SW7 2AZ, United Kingdom}

\author{Simon C. Webster}
\affiliation{Blackett Laboratory, Imperial College London, London SW7 2AZ, United Kingdom}

\author{Jacopo Mosca Toba}
\affiliation{Blackett Laboratory, Imperial College London, London SW7 2AZ, United Kingdom}

\author{Ollie Corfield}
\affiliation{Blackett Laboratory, Imperial College London, London SW7 2AZ, United Kingdom}

\author{George Porter}
\affiliation{Blackett Laboratory, Imperial College London, London SW7 2AZ, United Kingdom}

\author{Richard C. Thompson}
\affiliation{Blackett Laboratory, Imperial College London, London SW7 2AZ, United Kingdom}

\date{\today}

\begin{abstract}
{Measurement-based cooling is a method by which a quantum system, initially in a thermal state, can be prepared  {probabalistically} in its ground state through some sort of measurement. } This is done by making a measurement that heralds the system being in the desired state.  Here we demonstrate the application of a measurement-based cooling technique to a trapped atomic ion.  The ion is pre-cooled by Doppler laser cooling to a thermal state with a mean excitation of $\bar n \approx 18$ and the measurement-based cooling technique selects those occasions when the ion happens to be in the motional ground state.  The fidelity of the heralding process is greater than 95\%.  This technique could be applied to other systems that are not as amenable to laser cooling as trapped ions.

\end{abstract}

\maketitle

\section{Introduction}
\label{sec:introduction}

{Preparing a quantum system in  a pure state allows the system's quantum mechanical behaviour to be observed and exploited.} Doppler laser cooling, as first demonstrated in trapped ions, makes use of the repeated absorption and emission of photons, and the associated transfer of momentum, to cool quantum systems far below their initial thermal state \cite{eschner2003laser}. In order to reach the ground state, however, it is often necessary to use a sideband cooling technique after initial Doppler cooling \cite{Diedrich1989sidebandcooling,monroe1995demonstration}.  Laser cooling techniques have now been applied to a wide variety of systems, including atoms and ions \cite{wineland1978lasercooling,Neuhauser1978lasercooling_atom,Neuhauser1980laser_cooling_Ba+}, complex molecules \cite{Barry2012lasercoolingmolecules,Hummon2013lasercoolingmolecules,Shuman2009lasercoolingmolecules,shuman2010laserlasercoolingmolecules} and mechanical oscillators \cite{chan2011lasercooling_resonator,teufel2011lasercooling_resonator}. While laser cooling can be very successful for some systems, there exist many systems of interest for which direct application of laser cooling is either impractical or impossible, particularly if the intention is to cool the motion to the ground state. Doppler laser cooling requires a strong  transition to absorb photons from the laser beam  and return the system to the ground state by re-emission.  Sideband cooling additionally requires a weak transition to a metastable state \cite{Diedrich1989sidebandcooling}.  One or more of these may not exist or there may be some other impediment to effective cooling: for instance in a trapped molecule the additional complexity of the level structure can increase the number of photon scattering events required to reprepare the molecule by optical pumping during sideband cooling, an effect exacerbated if the trapping potential is state dependent, as each scattering event then has a higher probability of heating the molecule \cite{caldwell2020}.

{In a micromechanical oscillator, there {may be} no suitable energy level structure available for laser cooling, though direct measurement of the position and momentum quadratures using laser pulses can result in an effective temperature much less than that of the environment \cite{vanner2013}.  }   In a very different system also not amenable to laser cooling, the BASE collaboration aims to detect  proton spin-flips in a Penning trap.  In order to increase the fidelity of this detection, they first check that the proton's cyclotron energy  is sufficiently low, and if it is not, they allow it to rethermalise and then repeat the measurement \cite{mooser2013}. In this way, spin-flip detection is only carried out when the proton has a low cyclotron energy.

{In other cases an alternative method of cooling is necessary, especially if the requirement is to cool the system to its quantum mechanical ground state. Measurement-based cooling is one such method. In its simplest form, this relies on performing a measurement on a system whose result depends on its motional state. If the system starts in a thermal distribution of motional states there is a non-zero probability that it is in the ground state and so by performing a state-dependent measurement, we can select on the result of the measurement to obtain the subset of the thermal distribution which meets the requirement. After each measurement the system is allowed to rethermalise.   One specific measurement outcome heralds the system being in the ground state and all other data are discarded \cite{puebla2020measurementcooling,Montenegro2018measurementcooling,puebla2020measurementcooling,Bergenfeldt2009measurementcooling}.  Thus, measurement-based cooling enables us to probe quantum behaviour tailored to the motional ground state without the need for cooling all the population to that  state. {It is simpler to implement than optical sideband cooling, for example, but it has the disadvantage that because it is a probabilistic technique, not all attempts are successful.  It  can therefore  become inefficient when the mean excitation of the motional state is large. Such probabilistic state preparation methods have already been used to prepare atomic and molecular systems into a pure internal state, for instance to initialise molecular ion systems into specific rovibrational states. }\cite{vogelius2006,chou2017}.  {The work presented here is an extension of these techniques to the external motional states.}
}

One proposed cooling method considers an oscillator coupled to a two level system via a Jaynes-Cummings type Hamiltonian \cite{puebla2020measurementcooling}. By using a numerically optimised time-varying coupling, Puebla et al. \cite{puebla2020measurementcooling} show it is possible to perform a conditional logic operation on the two-level qubit-like system, such that the state of the qubit is unchanged if the ion is in the motional ground state, and it is flipped if in any other motional state. Here we propose a simplification of this scheme which makes use of rapid adiabatic passage to perform the logic operation, and  demonstrate the use of this scheme to probabilistically prepare a single trapped ion in the ground state of one of its harmonic oscillator modes with high fidelity.

\section{Method}

\subsection*{Measurement-based Cooling}

We consider a composite quantum system consisting of a qubit with states $\ket{g}$ and $\ket{e}$ and a harmonic oscillator mode with states $\ket{n}$, $n\geq0$. The qubit state must be measurable with high fidelity, and for at least one of the measurement outcomes no change of the harmonic oscillator state should result as a by-product of the measurement process. We assume that the two systems can be coupled via a controllable Jaynes-Cummings Hamiltonian:

\begin{equation}
    \label{eq:JC_Hamiltonian}
    H(t) = \frac{\hbar\Omega(t)}{2} \left(a\sigma_+ e^{i\delta(t)t} + a^\dagger\sigma_- e^{-i\delta(t)t}\right),
\end{equation}

\noindent where $\Omega(t)$ and $\delta(t)$ are a suitably defined time-varying Rabi frequency and detuning.

To perform the filtering operation, a conditional logic operation must take place, with the qubit selectively excited from its initial state $\ket{g}$ to $\ket{e}$ if and only if the motion is not in the state $\ket{0}$:
\begin{align}
    \ket{g}\ket{0} &\rightarrow \ket{g}\ket{0},\\
    \ket{g}\ket{n\neq0} &\rightarrow \ket{e}\ket{{n-1}}. \label{eq:RAP}
\end{align}
The qubit state is then measured, and a measurement outcome of $\ket{g}$ heralds the oscillator being in the desired ground state. If the oscillator is initially in a thermal state characterised by a mean excitation value $\bar{n}$, the thermal distribution of the motional states is given as

\begin{equation}
    \label{eqn:thermal_state}
    \rho_{th} = \sum_{n=0}^{\infty} p_{n} \ket{n}\bra{n} ,\quad p_{n} = \frac{\bar{n}^{n}}{(\bar{n}+1)^{n+1}}.     
\end{equation}
\noindent This indicates that the probability  the ion is in the motional ground state is $p_{0} = 1/(\bar{n}+1)$. In the case of a qubit measurement of $\ket{e}$, the oscillator can be re-thermalised and the process repeated until the desired heralding measurement result is obtained.

\begin{figure}
    \includegraphics[width=0.9\linewidth]{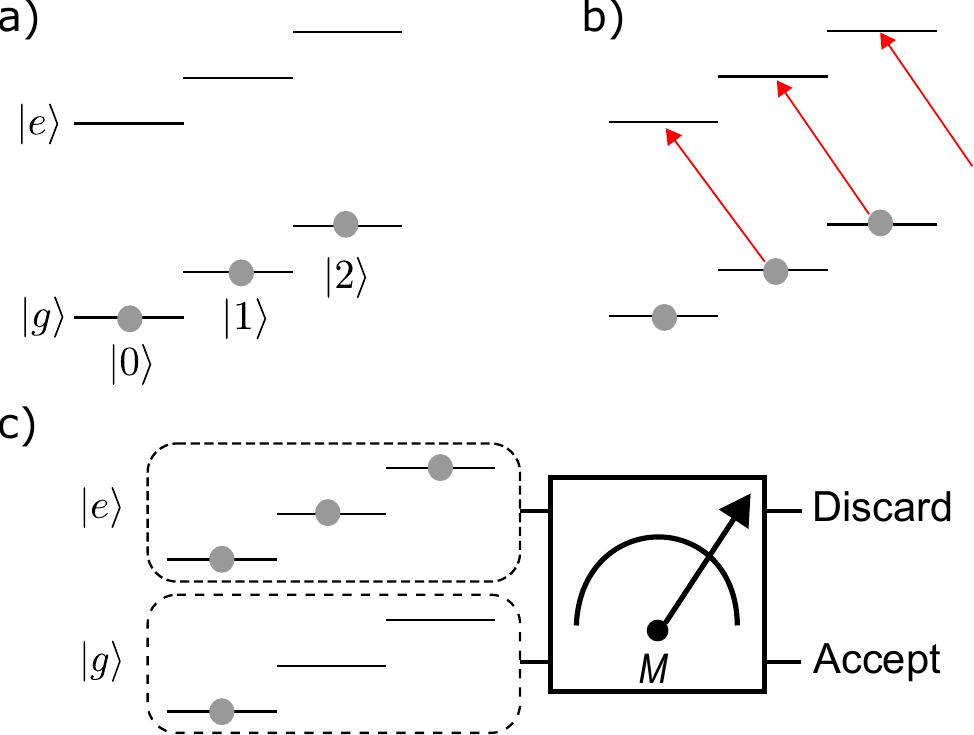}
    \centering
    \caption{Principle of measurement-based cooling. (a) Initial state after Doppler cooling. A composite system is initialised to its qubit ground state with thermal state $\ket{g}\bra{g} \otimes \rho_{\rm th}$ during the state-preparation period. (b) Population transfer by RAP applied to the red sideband. The system is excited to  $\ket{e}$ with its motional quantum number reduced by one unit for Fock states $\ket{n\ne 0}$  while the population in the motional ground state $\ket{0}$ remains in $\ket{g}$. Consequently, the qubit ground state is associated only with the motional ground state. (c) Projective measurement on the qubit states, $\ket{g}$ and $\ket{e}$. The outcome of the  measurement  is correlated with the occupied motional state. Thus, the motional ground state can be isolated from the rest of the motional states by selecting only those cases where the qubit state is measured to be the ground state.}
    \label{fig:cooling_scheme}
\end{figure}

This conditional operation, however, cannot be realised by a simple resonant pulse as the coupling strength of the transitions driven by the Hamiltonian of Eq. \ref{eq:JC_Hamiltonian} varies with the motional state $n$. In Puebla et al. \cite{puebla2020measurementcooling}, they considered the case where only the detuning $\delta(t)$ varies, while the coupling strength has a fixed value. They showed that by using a numerical optimisation technique, a suitable form of $\delta(t)$ can be constructed to perform the required conditional operation. A simpler method to perform the operation is to use a rapid adiabatic passage (RAP), since it is able to invert population between two internal states in a way that is insensitive to the coupling strength used \cite{allen1987optical}. Rapid adiabatic passage can be performed by varying the control parameters in relatively simple ways, and is robust against errors in the form of the control parameters. This robustness may also allow RAP based techniques to be used even in the case where the oscillator is anharmonic. 

The system is first prepared in the qubit ground state $\ket{g}$ with the motional states following a thermal distribution, as illustrated in Fig \ref{fig:cooling_scheme}(a).  Application of a RAP on the Jaynes-Cummings Hamiltonian Eq. \ref{eq:JC_Hamiltonian} brings all the population into the excited qubit state $\ket{e}$, except when the oscillator has motional quantum number $n = 0$.  This step  is illustrated in Fig.  \ref{fig:cooling_scheme}(b) and is 
described by Eq \ref{eq:RAP} above.

This process leaves the motional ground state untouched while other motional states are mapped onto a different internal state, so that qubit measurement then provides the required heralding signal to indicate a ground state cooled oscillator (Fig \ref{fig:cooling_scheme}(c)).

An imperfect RAP process degrades the cooling fidelity, that is to say the conditional probability of the motion being in the ground state given a measurement of the qubit as being in $\ket{g}$, since it can leave a motionally excited ion in the state $\ket{g}$. Using a simplified model where we assume that the probability of a successful RAP is independent of motional state $n$, and characterised by a transfer failure probability $\epsilon$, then the conditional success probability is given by

\begin{equation}
    \label{eq:RAP_error}
    p(n=0|g) = \frac{1}{1+\epsilon\bar{n}}.
\end{equation} 

If for a given $\epsilon$ and $\bar{n}$ this probability is lower than required then the RAP-measurement cycle can be repeated multiple times, with the operation considered successful if and only if each measurement yields a result of $\ket{g}$. For $m$ cycles, 

\begin{equation}
    \label{eq:RAP_error2}
    p(n=0|\ m\times g) = \frac{1}{1+\epsilon^m\bar{n}}.
\end{equation} 

Each cycle reduces the probability that a measurement result of $\ket{g}$ is a false positive result.

\section{Implementation}
\label{sec:Experiment}
\subsection*{Ion trap}
We demonstrate the method described above using a single trapped ion. Trapped ions are an ideal system in which to first demonstrate techniques such as this due to the high degree of control possible in the system, and the degree to which they can be isolated from the environment. We will describe the ion trap system before discussing the modifications to the generic measurement-based cooling method described above that are required for the demonstration.

\begin{figure}
    \includegraphics{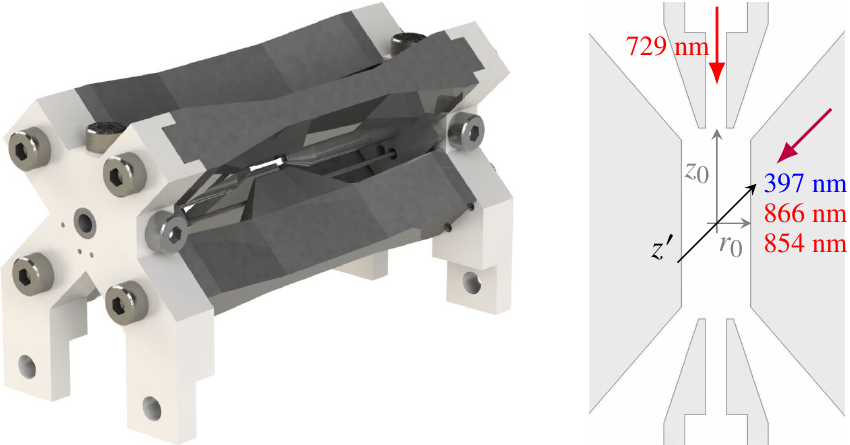}
    \caption{
    Linear ion trap (left), based on an Innsbruck-style blade trap design detailed in~\cite{Gulde2003}.
    The drawing (right) shows a cross section through the  electrodes.
    The trap dimensions are $z_0 = \SI{2.75}{\mm}$ and $r_0 = \SI{1}{\mm}$.
    Also shown is the direction of the lasers relative to the magnetic field quantisation axis, labelled $z'$.
    }
    \label{fig:rf}
\end{figure}

{The experiment described in this study has been carried out in a macroscopic linear RF trap \cite{corfield2021}} (see Fig.~\ref{fig:rf} for details of the electrode structure). The trap employs a combination of static and time-varying electric fields to achieve three-dimensional confinement of $^{40}$Ca$^{+}$ ions. The trap parameters are set such that the radial motional frequency ($\approx \SI{1.81}{\mega\hertz}$) is  higher than the axial one ($\approx \SI{1.06}{\mega\hertz}$). The motional states considered are those associated with the axial motion.

\begin{figure}    
    \includegraphics{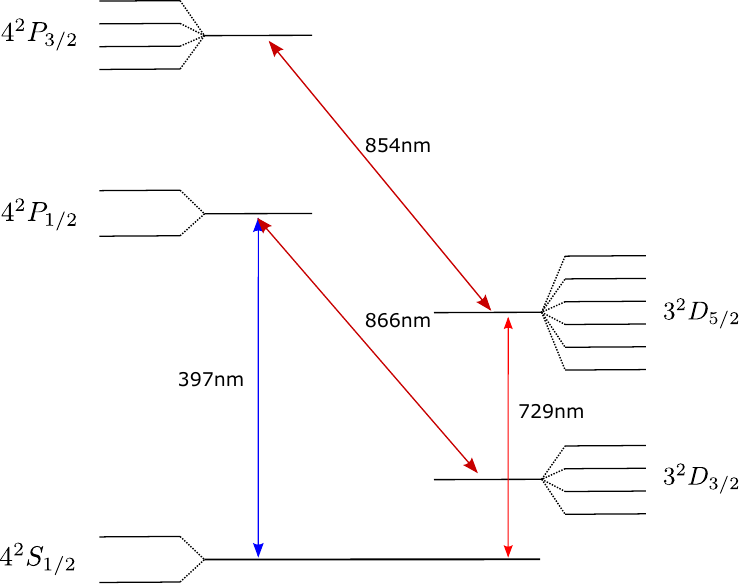}
    \centering
    \caption{Atomic structure and relevant transitions of$\;^{40}$Ca$^{+}$.  Laser cooling is carried out using the lasers at 397 and 866 nm.  The laser at 729 nm is used to coherently address the qubit transition, and the 854 nm laser is used to empty the D$_{5/2}$ state at the end of a measurement.}
    \label{fig:ca_diagram}
\end{figure}

The quantum system represented by a trapped ion effectively consists of an internal atomic  state, which is simplified to a two-level system, and a motional state that is set by a harmonic trapping potential. In our experiment, the two-level qubit states are defined by $\mathrm{S}_{1/2,m_j=1/2}$ ($\ket{g}$) and $\mathrm{D}_{5/2,m_j=1/2}$ ($\ket{e}$), and a quadrupole $\pi$-transition between the two is used to coherently manipulate the qubit state (see Fig. \ref{fig:ca_diagram}). The $\pi$ transition is addressed with a narrow-linewidth \SI{729}{\nano\m} laser. This laser is locked to an external high-finesse cavity, and its frequency is stabilised using the Pound–Drever–Hall (PDH) technique to achieve a laser linewidth narrower than \SI{400}{\hertz}, which is a critical requirement for laser-induced coherent operations.

As shown in Fig. \ref{fig:rf}, the main spectroscopic laser at \SI{729}{\nano\m} is directed along the trap axis. This configuration maximises the Lamb-Dicke (LD) parameter $\eta$ associated with the axial motion and nulls $\eta$ for the radial motions. The other lasers  are directed along the quantisation axis, which is defined by a weak external magnetic field  of $\approx 0.32$ mT produced by a coil placed outside the vacuum chamber. A linearly polarised \SI{397}{\nano\m} laser addresses the dipole transition $\mathrm{S}_{1/2} \longleftrightarrow \mathrm{P}_{1/2}$ and is used for Doppler cooling. The decay on this transition is used to discriminate between the two qubit states -- if the ion is in the internal ground state $\ket{g}$, the Doppler beam repeatedly excites the ion to the short-lived $\mathrm{P}_{1/2}$ state from where it decays back to the $\mathrm{S}_{1/2}$ state, emitting a photon; otherwise the ion is decoupled from the laser, and no photons are emitted. Emitted photons are detected by a photomultiplier tube (PMT), and the resulting photon distribution is Poissonian with different mean photon numbers depending on the projected qubit state. State-dependent fluorescence therefore allows for discrimination between $\mathrm{S}_{1/2} \ (\ket{g})$ and $\mathrm{D}_{5/2} \ (\ket{e})$.  This measurement takes about 1.5 ms and has a fidelity of \SI{99.5}{\percent}.  This measurement process only leaves the motional state unaffected if the system is in the excited qubit state $\ket{e}$.  (Note that this requires a small change to the scheme, {which is illustrated in Fig.} \ref{fig:cooling_actual_scheme} and discussed below.)

The  undesired metastable state $\mathrm{D}_{3/2}$  can be occupied via spontaneous decay from $\mathrm{P}_{1/2}$ and so this population is pumped back to $\mathrm{S}_{1/2}$ via the dipole  transition at \SI{866}{\nano\m}  during laser cooling.  Any population left in $\mathrm{D}_{5/2}$ from coherent operations at  \SI{729}{\nano\m} is pumped back to $\mathrm{S}_{1/2}$ via the dipole transition at  \SI{854}{\nano\m} during the state preparation.  

\subsection*{Applying the method to a trapped ion}

\begin{figure}
    \includegraphics[width=0.9\linewidth]{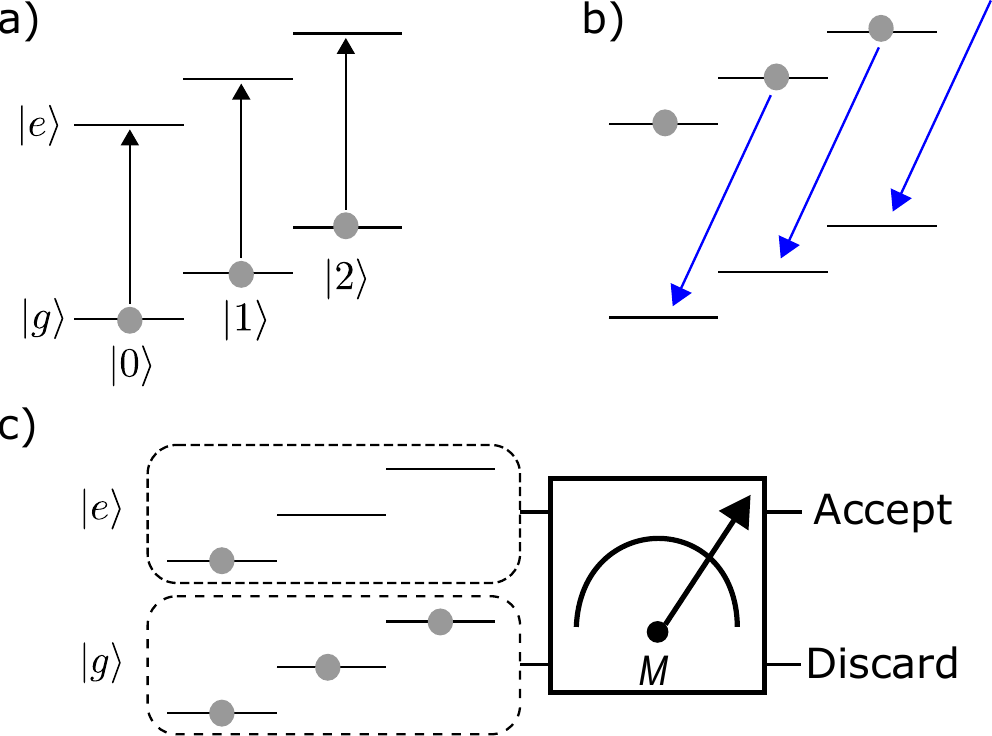}
    \centering
    \caption{{Actual sequence of processes to implement measurement-based cooling in our experiments. (a) RAP applied to the carrier. The internal state of the ion is initialised to the excited state $\ket{e}$ with its motional distribution unaltered. (b) RAP applied to the blue sideband. The ion is de-excited back to $\ket{g}$ for $n\neq0$ so that the ion is left in $\ket{e}$ if and only if its associated motional state is $\ket{0}$. (c) Projective measurement on $\ket{g}$ and $\ket{e}$. The motional ground state is heralded by the measurement.}}
    \label{fig:cooling_actual_scheme}
\end{figure}

{The interaction between the trapped ion and the radiation is described by the Hamiltonian}
\begin{equation}
    \label{hamiltonian}
    H=(\hbar/2)\Omega_0\sigma_+\exp{[i\eta(\hat{a}e^{-i\nu t}+\hat{a}^{\dagger}e^{i\nu t})]} + \text{H.c.}
\end{equation} 

{This equation can be solved to find the coupling strength between  state $|g\rangle|n\rangle$ and $|e\rangle|n+s\rangle$, characterised by the Rabi frequency $\Omega_{n,n+s}$} \cite{Leibfried2003}.


  When the ion is in the Lamb-Dicke regime $(\eta\ll1)$ \cite{wineland1988}, the Hamiltonian Eq.\ref{hamiltonian} reduces to the  Jaynes-Cummings Hamiltonian Eq. \ref{eq:JC_Hamiltonian}  {when driving a first-order sideband transition  ($s=-1$).  In these circumstances the expressions for the coupling strengths on the first-order sideband and carrier 
     reduce to a simple form} \cite{Leibfried2003}
  \begin{equation}
    \label{rabi-frequencies-2}
    \Omega_{n,n-1}=\Omega_0 \eta\sqrt{n}\ \ \ \text{and}\ \ \ \Omega_{n,n}=\Omega_0.
\end{equation}  
    
    After Doppler cooling, the ion is typically not in the Lamb-Dicke regime, however the only result of this is that the coupling strengths in Eq. \ref{rabi-frequencies-2} are modified for larger values of $n$, and since the RAP process is insensitive to coupling strength, this has no effect on the process in principle. 

The measurement process used in the trapped ion does however require a modification to the cooling protocol 
 (see Fig. \ref{fig:cooling_actual_scheme}). Since the initially prepared $\ket{g}$ state is indicated by the emission of photons by the ion, this will lead to heating of the oscillator mode. Conversely, $\ket{e}$ is indicated by the lack of emitted photons, leading to no heating. This means that the conditioning measurement should be detection of the qubit as being in $\ket{e}$. This requires that the qubit be initialised in $\ket{e}$ and an anti-Jaynes-Cummings interaction, which is the blue sideband transition, used for the conditional logic operation. Since outside the Lamb-Dicke regime the qubit carrier transition has a weak dependence on motional quantum number $n$, a RAP on the carrier can be used to transfer the qubit state from $\ket{g}$ to $\ket{e}$ in a motionally insensitive way. 

\subsection*{Rapid Adiabatic Passage}

Rapid adiabatic passage has been applied to many different two level quantum systems, including trapped ions \cite{Watanabe2011RAPsideband,WunderlichChr2007RAPiontrap,Toyoda2011RAP_on_SB}. While an ion coupled to a motional mode is not a two-level system, we operate in a parameter regime where only a single sideband is addressed and multi-level effects are negligible.

The RAP implementation used involves applying a frequency chirp to the driving laser field while the power of the laser is smoothly ramped up and down to avoid diabatic processes which can occur if the laser is abruptly switched on and off. There are many possible ways to perform these modulations of the control parameters; in our experiment we use a simple linear frequency sweep, while the amplitude modulation follows a squared-sine envelope:

\begin{align}
    \label{eqn:RAP_pulse}
    \delta(t) &= \frac{\delta_{0}}{T}\left(t-\frac{T}{2}\right),\\
    \Omega(t) &= \Omega_{\rm peak}\sin^{2}\left(\frac{\pi t}{T}\right),  \quad 0 \leq t \leq T
\end{align}
\noindent where $T$ is the total duration of the RAP pulse, $\delta_0$ is the range of the detuning chirp, and $\Omega_{\rm peak}$ the peak Rabi frequency. 

RAP can give almost unity population transfer efficiency even when the ion is only Doppler cooled, where many different motional states are statistically occupied; this means that population transfer between two internal states can be carried out for different $n$ using the same laser parameters, which is crucial for the measurement-based cooling method presented here.  

\section{Results}
\label{sec:Results}

\begin{figure}
    \centering
    \includegraphics[width=1.0\linewidth]{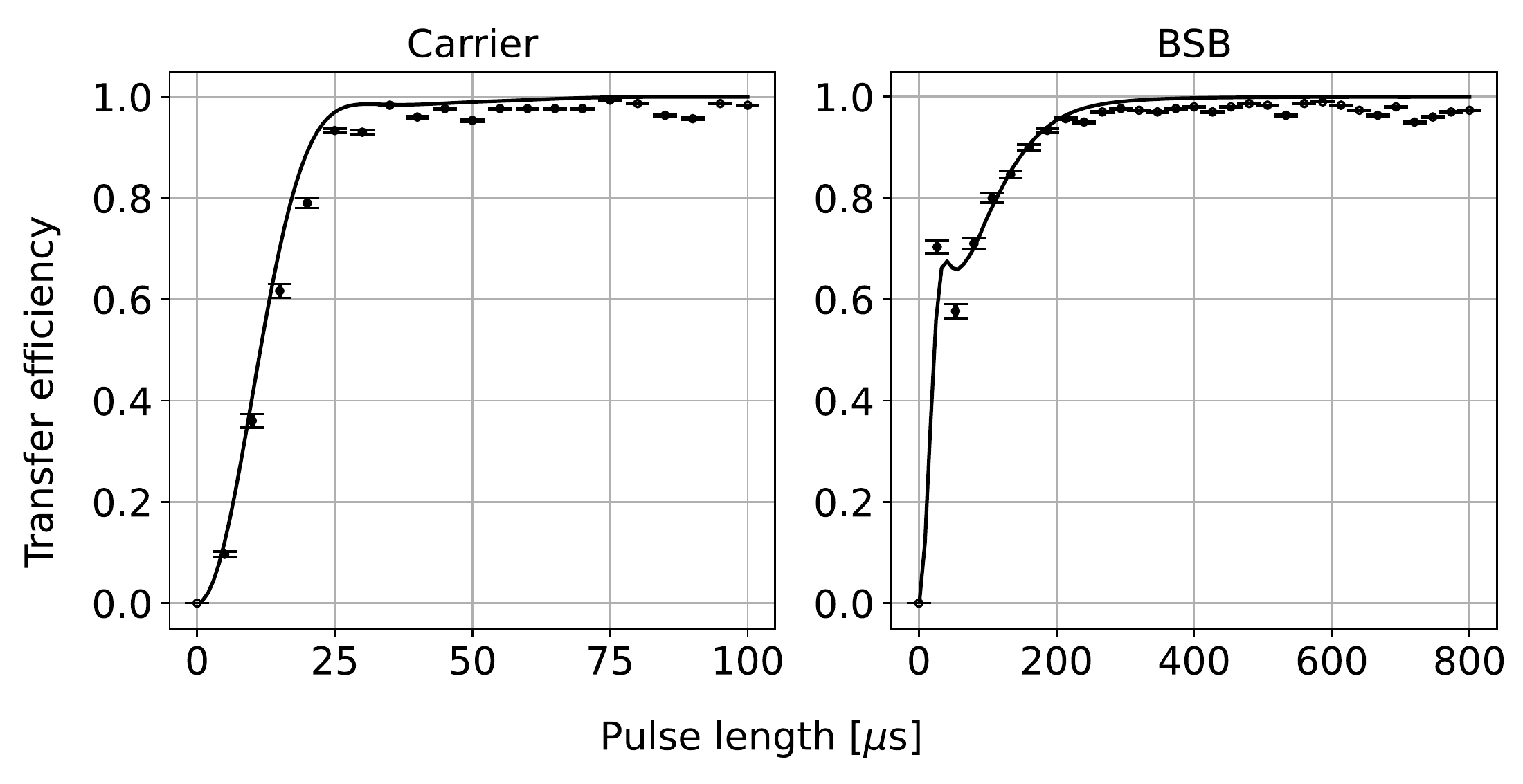}
    \caption{Transfer efficiency of RAP as a function of total duration of RAP pulse $T$, which includes ramping up and down periods. RAP operations are applied to the carrier (left) and the blue-sideband transition (right) after the ion is only Doppler cooled. The detunings of the laser are linearly scanned with a total detuning range of 200\,kHz (carrier) and 40\,kHz (blue sideband). In all scans, the power of the laser is set such that it gives {peak} carrier Rabi frequency {$\Omega_{\rm peak}$} of 83\,kHz. The simulation  results are computed by numerically solving the relevant quantum master equation that accounts for the frequency and amplitude modulation of the driving light field. The transfer efficiency of RAP on the blue sideband initially increases with total length of the pulse $T$, but there is a small dip around $50\,$\textmu s. This is seen both in the numerical calculation and  in the experimental results. This feature is attributed to breakdown of adiabaticity during the RAP process \cite{Lu2007non_adiabatic_transition}. }
    \label{fig:transfer_efficiency}
\end{figure}

The performance of measurement-based cooling is heavily dependent on the fidelity of population transfer by RAP operations. In Fig. \ref{fig:transfer_efficiency}, the transfer efficiencies of RAP on carrier and blue-sideband transitions are presented as a function of the total duration of the driving pulses for a fixed frequency chirp and peak Rabi frequency. {The simulation  shows that the transfer efficiency converges to unity as the pulse length $T$ becomes longer because the dynamics of the process become more adiabatic. Although experimental data also indicates that the transfer efficiency approaches a saturation value close to unity, the transfer efficiency still fluctuates and saturates at a value of around 95$\%$ even when the pulse time is much longer than the pulse time for which it should be very close to unity according to  the simulation curve. Thus, we adopt the shortest pulse time that lies on the flat region of the simulation and experimental curves.  In modelling the  measurement-based cooling scheme we assume a  transfer efficiency of  95\%.} For the carrier transition $T=$ 35\,\textmu s, $\Omega_{\rm peak}/2\pi=83$\,kHz and $\delta_0/2\pi=200 $\,kHz while for the sideband transition $T=$ 250\,\textmu s, $\Omega_{\rm peak}/2\pi=5.8$\,kHz and $\delta_0/2\pi=40$\,kHz. {In order to set the frequency sweep range properly, we perform numerical simulations beforehand and ensure the sweep range is just big enough to avoid breakdown of the adiabaticity condition. Note that the transfer efficiency on the blue sideband is roughly the same as for the carrier even though the strengths of the sideband transitions depend strongly on $n$.}

\begin{figure*}
    \includegraphics[width = 1\linewidth]{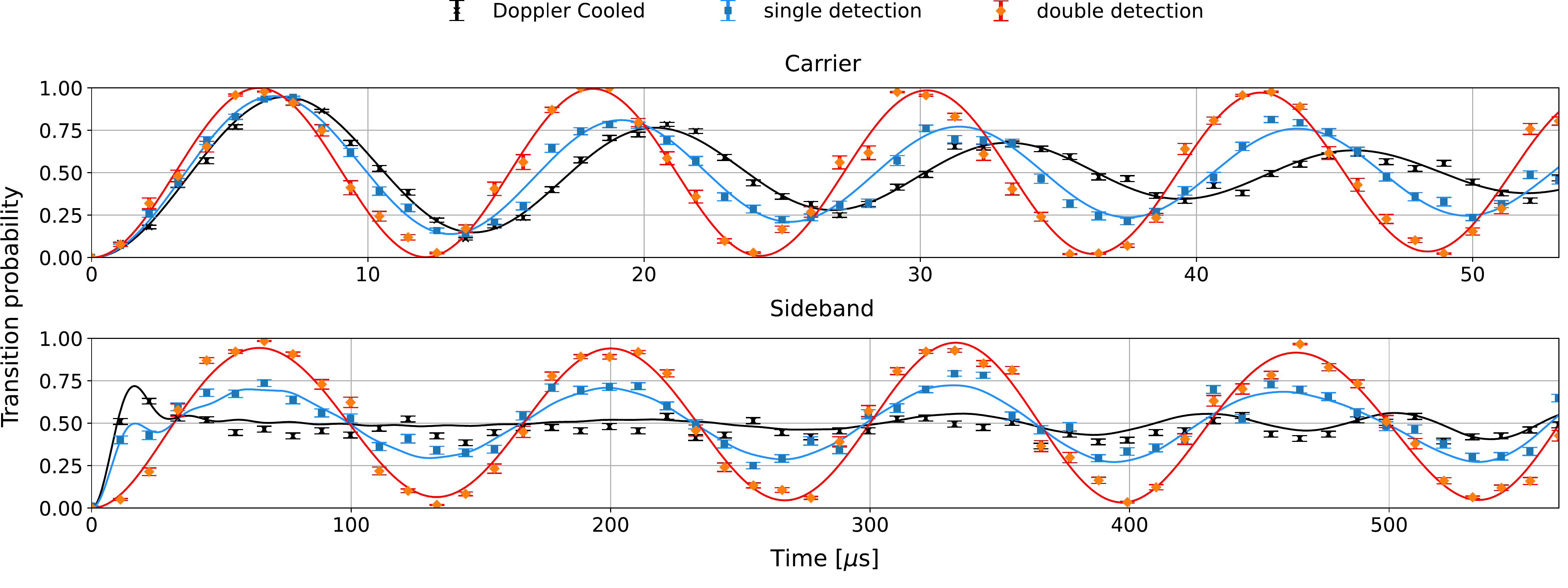}
    \caption{Rabi oscillations on the carrier (upper pane) and first-order sideband (lower pane) with and without measurement-based cooling. The black curves correspond to data obtained using Doppler cooling only; the blue curves correspond to  a single cycle of measurement-based cooling; and the orange curves correspond to two cycles of cooling. See the text for more details. The solid lines represent fit curves for each scan to the model described by Eqn.\ref{eqn:rabi_mixed_state}.}
    \label{fig:cooling_time_scan}
\end{figure*}

To evaluate the success of the cooling scheme, we make measurements of resonant Rabi flopping on both the carrier and first-order sideband transitions. For this measurement we measure the state of the ion qubit after a pulse of 729-nm light of variable duration. In figure \ref{fig:cooling_time_scan}, the black traces show the Rabi oscillations on these two transitions for the Doppler cooled ion initialised in state $\ket{g}$. The relatively weak dependence of carrier Rabi frequency on motional state results in a slow decay in the oscillation, while for the blue sideband, the much stronger motional state dependence results in a rapid equalisation of the population in the internal states.  A fit to the carrier data suggests $\bar{n} \approx 18$; this means the ion is in its motional ground state roughly 5 percent of the time, {assuming the ion is initialised to a thermal state} after Doppler cooling. 

The blue curves show the Rabi oscillations obtained after performing one cycle of measurement-based cooling. To perform this single cycle of cooling, after the ion is Doppler cooled and prepared in $\ket{g}$ the carrier RAP is used to move the ion into $\ket{e}$ before the sideband RAP performs the conditional logic operation. A heralding state measurement is performed, {and the data is only used if the measurement implies that the system is in the motional ground state.  Following this,}  the Rabi oscillation on either the carrier or red sideband  is measured. This process is repeated 900 times for each probe pulse length,  and a heralded success is found with a probability of 12\% (averaged over the complete dataset). 

This heralded success probability is much larger than the estimated initial motional ground state population before measurement-based cooling is applied (approximately 5\%). {This indicates that selecting data conditioned on the ion being in $\ket{e}$ during the cooling process does not herald the ground state exclusively, due to imperfect RAP. This can be seen by looking at the Rabi oscillation observed on the {red} sideband:  a clear oscillation is observed at the frequency expected for an ion in the motional ground state, however the amplitude of oscillation indicates that a large fraction of the motional state population still lies outside the ground state.}

The Rabi oscillations on the carrier still decay after one cycle of RAP, but at a slower rate than in the Doppler-cooled case. Note that the oscillation frequency rises slightly after application of RAP, because the Rabi frequency is higher for $n=0$ than for excited motional states.

In our simple model where the excited motional state population is caused by a motional-state independent RAP failure, the probability of being in the motional ground state after a {single} successful heralding event is given by a modification of Eq. \eqref{eq:RAP_error}: 
\begin{equation}
	\label{eqn:RAP_cooling_0}
		p(n = 0 | e) =\frac{1}{1+2\epsilon\bar{n}} 
\end{equation}
{where the extra factor of 2 arises from the addition of another RAP step on the carrier (assuming, for simplicity, that the transfer efficiency for the carrier is equal to that for the sideband).  The} probabilities of being in the other motional states are given by
\begin{equation}
	\label{eqn:RAP_cooling}
		p(n \neq 0 | e) = \frac{\epsilon}{1+2\epsilon\bar{n}}\left( \frac{\bar{n}}{1+\bar{n}} \right)^n.
\end{equation}

The excitation probability after the probe resonant pulse for heralded events is then given by
\begin{equation}
	\label{eqn:rabi_mixed_state}
	P(t) = \sum_{n=0} p(n | e) \sin^{2}\left( \frac{\omega_n t}{2} \right)
\end{equation}
where the oscillation frequencies depend on the motional-state dependent Rabi frequencies, $\omega_n=\Omega_{n,n}$ for the carrier transition, and $\omega_n=\Omega_{n,n+1}$ for the sideband transition, $\Omega_{n,n+s}=\Omega_{n+s,n}$ being the Rabi frequency for the  transition between motional states $n$ and $n+s$ in Eq. {\ref{rabi-frequencies-2} \cite{Wineland1979state_dependent_rabi}. Assuming we know the Lamb-Dicke parameter $\eta$, we know the state-dependent Rabi frequencies for all $n$, and Eq. \ref{eqn:rabi_mixed_state} can be parameterised by $\bar{n}$ and ${\eta}$. We allow the motional ground state population $p_0=p(0|e)$ and ${\eta}$ to float in our fit, giving an estimation of $p_0$. The population outside the ground state is modeled as a thermal distribution for simplicity.   The fit to the Rabi oscillation on the sideband with the single stage of RAP  gives $p_0 \approx 0.49$. This is significantly improved from the initial ground state population $p_0\approx 0.05$, but is not yet close enough to unity to be useful. The value of $\bar n$ for the fitted thermal distribution comes out to be close to the original thermal value, as expected.

The success probability for this measurement-based cooling scheme can be increased by performing {additional cycles of sideband RAP-measurement immediately after the first cycle, and conditioning on all measurements finding the ion in $\ket{e}$}. 

{The probability of being in the motional  state after $m$ cycles is now}

\begin{equation}
	\label{eqn:RAP_cooling_0_m}
		p(n = 0 | m \times e) =\frac{1}{1+2\epsilon^m\bar{n}}. 
\end{equation}

The orange curves in Fig. \ref{fig:cooling_time_scan} show the oscillations for double measurement ($m$=2) heralding, {where 5\% of the complete data set is accepted in the post-processing}. Here, the data is consistent with the ion being in the ground state with high probability after this double measurement heralding. The oscillations on both the carrier and the red sideband have high visibility and slow decay, as expected.  The fit to the sideband Rabi oscillation indicates that $p_0$ is increased to $\approx 0.96$.

\section{Discussion}
\label{Discussion}
The measurement-based cooling scheme presented here consists of two steps -- one is to map the motional ground state and any excited motional states onto two different internal states using a sequence of RAP operations, and the other is to make a projective measurement and discard instances where the result of the measurement is the ground state $\ket{g}$.  The method's cooling efficiency therefore relies significantly on the transfer efficiency of RAP and state detection fidelity.

As presented in Sec. \ref{sec:Results}, the ground state population $p_{0}$ after a single cycle of RAP and the state detection is still low ($\approx 0.49$). {This low $p_{0}$ is attributed to imperfect RAP operations. Using our assumed transfer failure probability of $\epsilon=0.05$,} Eq. \eqref{eqn:RAP_cooling_0} {estimates $p_{0} \approx 0.56$, which is somewhat higher than the observed value.} 
The additional cooling sequence, which consists of another RAP on the blue sideband and subsequent projective measurement, raises $p_{0}$, as expected from Eq. \eqref{eq:RAP_error2}. {However, this new value of  $p_{0}\approx 0.96$ is also  lower than the value expected from} Eq. \eqref{eqn:RAP_cooling_0_m}, {which is $\approx 0.98$.} Furthermore, we do not observe any further improvements on $p_{0}$ when we apply any more cooling sequences.  

These observations lead us to consider other sources of error.    
The projective measurement takes \SI{1.5}{\milli\s} to ensure a detection fidelity greater than \SI{99.5}{\percent}. The detection fidelity is high enough to avoid false positive detection of the motional ground state, but the time taken for the detection is long enough that it is necessary to consider heating during the detection time. The heating rate of our trap was measured to be approximately \SI{37}{phonons\per\second}. This rate implies a 5\% chance that an ion in the {motional} ground state after the {last} sideband RAP pulse will no longer be in the {motional} ground state by the conclusion of the measurement, {which leads to the increase in $p_{0}$. Moreover, the effect of the heating cannot be alleviated by applying more RAP operations since it occurs during the measurement, which is performed after the last RAP operation. So  after our measurement-based cooling sequence, $p_{0}$ saturates to a value below  unity no matter how many RAP operations are involved in the cooling process, and the measured heating rate is sufficient to account for the population outside the motional ground estimated from the fitting curve shown in Fig. \ref{fig:cooling_time_scan}.

\section{Conclusion}
\label{Conclusion}
We have experimentally applied a measurement-based cooling scheme to a single trapped ion. The ion is initially prepared in a thermal state with $\bar{n} \approx 18$ using Doppler cooling, which leaves most of the population outside the motional ground state. We employ a rapid adiabatic passage to separately map the populations of the motional ground state and the rest of the motional states onto the ion's internal excited state $\ket{e}$ and ground state $\ket{g}$ respectively. { A measurement on the internal degree of freedom is then used to herald the motional ground state. Imperfections in the rapid adiabatic passage limit the population in the motional ground state for these heralded instances, however by repeating the mapping-measurement procedure, the probability to find the system in the motional ground state is increased.
}
In this study, the cooling is performed on a single motional mode. In principle the scheme could be used to prepare a system in the ground state of multiple modes simultaneously by applying it sequentially to each mode. In practice however the joint probability of finding all the modes initially in the ground state, for most realistic scenarios, would rapidly become extremely small as more modes were added.

The technique has been demonstrated here in a trapped ion system in which other ground-state cooling methods could be applied, however, the scheme should have wider applicability. Ground-state cooling is very difficult for trapped molecules, especially where optical trapping leads to different potentials for different molecular states, but it has been proposed to use a technique similar to that discussed here for single molecules \cite{caldwell2020}.  For micro-mechanical resonators, existing cooling methods are either hard or impossible to implement.    {While this technique requires a Jaynes-Cummings type Hamiltonian to exist in the system, suitable couplings have already been demonstrated in micro-mechanical oscillator systems, for instance by embedding a nitrogen-vacancy center into a resonator and placing the combined system in a strong magnetic-field gradient to provide spin-motion coupling} \cite{arcizet2011}. 
 Being able to prepare a ground-state sample in these systems   will allow quantum effects, which would otherwise be washed out by being in a thermal state, to be observed.

\begin{acknowledgments}
We are very grateful to Mr Brian Willey for his technical support. { We thank M. R. Vanner and J. A.  Devlin for helpful comments on the manuscript. }
Financial support by EPSRC through all of \textit{Optimal control for robust ion trap quantum logic} Grant No.\ EP/P024890/1,
\textit{Hub in Quantum Computing and Simulation} Grant No. EP/T001062/1,
the \textit{Centre for Doctoral Training in Controlled Quantum Dynamics} Grant No.\ EP/L016524/1,
and the \textit{Training and Skills Hub in Quantum Systems Engineering} Grant No.\ EP/P510257/1 is gratefully acknowledged.
\end{acknowledgments}

\bibliography{bib}

\begin{thebibliography}{31}%
\makeatletter
\providecommand \@ifxundefined [1]{%
 \@ifx{#1\undefined}
}%
\providecommand \@ifnum [1]{%
 \ifnum #1\expandafter \@firstoftwo
 \else \expandafter \@secondoftwo
 \fi
}%
\providecommand \@ifx [1]{%
 \ifx #1\expandafter \@firstoftwo
 \else \expandafter \@secondoftwo
 \fi
}%
\providecommand \natexlab [1]{#1}%
\providecommand \enquote  [1]{``#1''}%
\providecommand \bibnamefont  [1]{#1}%
\providecommand \bibfnamefont [1]{#1}%
\providecommand \citenamefont [1]{#1}%
\providecommand \href@noop [0]{\@secondoftwo}%
\providecommand \href [0]{\begingroup \@sanitize@url \@href}%
\providecommand \@href[1]{\@@startlink{#1}\@@href}%
\providecommand \@@href[1]{\endgroup#1\@@endlink}%
\providecommand \@sanitize@url [0]{\catcode `\\12\catcode `\$12\catcode
  `\&12\catcode `\#12\catcode `\^12\catcode `\_12\catcode `\%12\relax}%
\providecommand \@@startlink[1]{}%
\providecommand \@@endlink[0]{}%
\providecommand \url  [0]{\begingroup\@sanitize@url \@url }%
\providecommand \@url [1]{\endgroup\@href {#1}{\urlprefix }}%
\providecommand \urlprefix  [0]{URL }%
\providecommand \Eprint [0]{\href }%
\providecommand \doibase [0]{https://doi.org/}%
\providecommand \selectlanguage [0]{\@gobble}%
\providecommand \bibinfo  [0]{\@secondoftwo}%
\providecommand \bibfield  [0]{\@secondoftwo}%
\providecommand \translation [1]{[#1]}%
\providecommand \BibitemOpen [0]{}%
\providecommand \bibitemStop [0]{}%
\providecommand \bibitemNoStop [0]{.\EOS\space}%
\providecommand \EOS [0]{\spacefactor3000\relax}%
\providecommand \BibitemShut  [1]{\csname bibitem#1\endcsname}%
\let\auto@bib@innerbib\@empty
\bibitem [{\citenamefont {Eschner}\ \emph {et~al.}(2003)\citenamefont
  {Eschner}, \citenamefont {Morigi}, \citenamefont {Schmidt-Kaler},\ and\
  \citenamefont {Blatt}}]{eschner2003laser}%
  \BibitemOpen
  \bibfield  {author} {\bibinfo {author} {\bibfnamefont {J.}~\bibnamefont
  {Eschner}}, \bibinfo {author} {\bibfnamefont {G.}~\bibnamefont {Morigi}},
  \bibinfo {author} {\bibfnamefont {F.}~\bibnamefont {Schmidt-Kaler}},\ and\
  \bibinfo {author} {\bibfnamefont {R.}~\bibnamefont {Blatt}},\ }\href
  {https://doi.org/10.1364/JOSAB.20.001003} {\bibfield  {journal} {\bibinfo
  {journal} {JOSA B}\ }\textbf {\bibinfo {volume} {20}},\ \bibinfo {pages}
  {1003} (\bibinfo {year} {2003})}\BibitemShut {NoStop}%
\bibitem [{\citenamefont {Diedrich}\ \emph {et~al.}(1989)\citenamefont
  {Diedrich}, \citenamefont {Bergquist}, \citenamefont {Itano},\ and\
  \citenamefont {Wineland}}]{Diedrich1989sidebandcooling}%
  \BibitemOpen
  \bibfield  {author} {\bibinfo {author} {\bibfnamefont {F.}~\bibnamefont
  {Diedrich}}, \bibinfo {author} {\bibfnamefont {J.~C.}\ \bibnamefont
  {Bergquist}}, \bibinfo {author} {\bibfnamefont {W.~M.}\ \bibnamefont
  {Itano}},\ and\ \bibinfo {author} {\bibfnamefont {D.~J.}\ \bibnamefont
  {Wineland}},\ }\href {https://doi.org/10.1103/PhysRevLett.62.403} {\bibfield
  {journal} {\bibinfo  {journal} {Phys. Rev. Lett.}\ }\textbf {\bibinfo
  {volume} {62}},\ \bibinfo {pages} {403} (\bibinfo {year} {1989})}\BibitemShut
  {NoStop}%
\bibitem [{\citenamefont {Monroe}\ \emph {et~al.}(1995)\citenamefont {Monroe},
  \citenamefont {Meekhof}, \citenamefont {King}, \citenamefont {Itano},\ and\
  \citenamefont {Wineland}}]{monroe1995demonstration}%
  \BibitemOpen
  \bibfield  {author} {\bibinfo {author} {\bibfnamefont {C.}~\bibnamefont
  {Monroe}}, \bibinfo {author} {\bibfnamefont {D.~M.}\ \bibnamefont {Meekhof}},
  \bibinfo {author} {\bibfnamefont {B.~E.}\ \bibnamefont {King}}, \bibinfo
  {author} {\bibfnamefont {W.~M.}\ \bibnamefont {Itano}},\ and\ \bibinfo
  {author} {\bibfnamefont {D.~J.}\ \bibnamefont {Wineland}},\ }\href
  {https://doi.org/10.1103/PhysRevLett.75.4714} {\bibfield  {journal} {\bibinfo
   {journal} {Phys. Rev. Lett.}\ }\textbf {\bibinfo {volume} {75}},\ \bibinfo
  {pages} {4714} (\bibinfo {year} {1995})}\BibitemShut {NoStop}%
\bibitem [{\citenamefont {Wineland}\ \emph {et~al.}(1978)\citenamefont
  {Wineland}, \citenamefont {Drullinger},\ and\ \citenamefont
  {Walls}}]{wineland1978lasercooling}%
  \BibitemOpen
  \bibfield  {author} {\bibinfo {author} {\bibfnamefont {D.~J.}\ \bibnamefont
  {Wineland}}, \bibinfo {author} {\bibfnamefont {R.~E.}\ \bibnamefont
  {Drullinger}},\ and\ \bibinfo {author} {\bibfnamefont {F.~L.}\ \bibnamefont
  {Walls}},\ }\href {https://doi.org/10.1103/PhysRevLett.40.1639} {\bibfield
  {journal} {\bibinfo  {journal} {Phys. Rev. Lett.}\ }\textbf {\bibinfo
  {volume} {40}},\ \bibinfo {pages} {1639} (\bibinfo {year}
  {1978})}\BibitemShut {NoStop}%
\bibitem [{\citenamefont {Neuhauser}\ \emph {et~al.}(1978)\citenamefont
  {Neuhauser}, \citenamefont {Hohenstatt}, \citenamefont {Toschek},\ and\
  \citenamefont {Dehmelt}}]{Neuhauser1978lasercooling_atom}%
  \BibitemOpen
  \bibfield  {author} {\bibinfo {author} {\bibfnamefont {W.}~\bibnamefont
  {Neuhauser}}, \bibinfo {author} {\bibfnamefont {M.}~\bibnamefont
  {Hohenstatt}}, \bibinfo {author} {\bibfnamefont {P.}~\bibnamefont
  {Toschek}},\ and\ \bibinfo {author} {\bibfnamefont {H.}~\bibnamefont
  {Dehmelt}},\ }\href {https://doi.org/10.1103/PhysRevLett.41.233} {\bibfield
  {journal} {\bibinfo  {journal} {Phys. Rev. Lett.}\ }\textbf {\bibinfo
  {volume} {41}},\ \bibinfo {pages} {233} (\bibinfo {year} {1978})}\BibitemShut
  {NoStop}%
\bibitem [{\citenamefont {Neuhauser}\ \emph {et~al.}(1980)\citenamefont
  {Neuhauser}, \citenamefont {Hohenstatt}, \citenamefont {Toschek},\ and\
  \citenamefont {Dehmelt}}]{Neuhauser1980laser_cooling_Ba+}%
  \BibitemOpen
  \bibfield  {author} {\bibinfo {author} {\bibfnamefont {W.}~\bibnamefont
  {Neuhauser}}, \bibinfo {author} {\bibfnamefont {M.}~\bibnamefont
  {Hohenstatt}}, \bibinfo {author} {\bibfnamefont {P.~E.}\ \bibnamefont
  {Toschek}},\ and\ \bibinfo {author} {\bibfnamefont {H.}~\bibnamefont
  {Dehmelt}},\ }\href {https://doi.org/10.1103/PhysRevA.22.1137} {\bibfield
  {journal} {\bibinfo  {journal} {Physical review. A, General physics}\
  }\textbf {\bibinfo {volume} {22}},\ \bibinfo {pages} {1137} (\bibinfo {year}
  {1980})}\BibitemShut {NoStop}%
\bibitem [{\citenamefont {Barry}\ \emph {et~al.}(2012)\citenamefont {Barry},
  \citenamefont {Shuman}, \citenamefont {Norrgard},\ and\ \citenamefont
  {DeMille}}]{Barry2012lasercoolingmolecules}%
  \BibitemOpen
  \bibfield  {author} {\bibinfo {author} {\bibfnamefont {J.~F.}\ \bibnamefont
  {Barry}}, \bibinfo {author} {\bibfnamefont {E.~S.}\ \bibnamefont {Shuman}},
  \bibinfo {author} {\bibfnamefont {E.~B.}\ \bibnamefont {Norrgard}},\ and\
  \bibinfo {author} {\bibfnamefont {D.}~\bibnamefont {DeMille}},\ }\href
  {https://doi.org/10.1103/PhysRevLett.108.103002} {\bibfield  {journal}
  {\bibinfo  {journal} {Phys. Rev. Lett.}\ }\textbf {\bibinfo {volume} {108}},\
  \bibinfo {pages} {103002} (\bibinfo {year} {2012})}\BibitemShut {NoStop}%
\bibitem [{\citenamefont {Hummon}\ \emph {et~al.}(2013)\citenamefont {Hummon},
  \citenamefont {Yeo}, \citenamefont {Stuhl}, \citenamefont {Collopy},
  \citenamefont {Xia},\ and\ \citenamefont
  {Ye}}]{Hummon2013lasercoolingmolecules}%
  \BibitemOpen
  \bibfield  {author} {\bibinfo {author} {\bibfnamefont {M.~T.}\ \bibnamefont
  {Hummon}}, \bibinfo {author} {\bibfnamefont {M.}~\bibnamefont {Yeo}},
  \bibinfo {author} {\bibfnamefont {B.~K.}\ \bibnamefont {Stuhl}}, \bibinfo
  {author} {\bibfnamefont {A.~L.}\ \bibnamefont {Collopy}}, \bibinfo {author}
  {\bibfnamefont {Y.}~\bibnamefont {Xia}},\ and\ \bibinfo {author}
  {\bibfnamefont {J.}~\bibnamefont {Ye}},\ }\href
  {https://doi.org/10.1103/PhysRevLett.110.143001} {\bibfield  {journal}
  {\bibinfo  {journal} {Phys. Rev. Lett.}\ }\textbf {\bibinfo {volume} {110}},\
  \bibinfo {pages} {143001} (\bibinfo {year} {2013})}\BibitemShut {NoStop}%
\bibitem [{\citenamefont {Shuman}\ \emph {et~al.}(2009)\citenamefont {Shuman},
  \citenamefont {Barry}, \citenamefont {Glenn},\ and\ \citenamefont
  {DeMille}}]{Shuman2009lasercoolingmolecules}%
  \BibitemOpen
  \bibfield  {author} {\bibinfo {author} {\bibfnamefont {E.~S.}\ \bibnamefont
  {Shuman}}, \bibinfo {author} {\bibfnamefont {J.~F.}\ \bibnamefont {Barry}},
  \bibinfo {author} {\bibfnamefont {D.~R.}\ \bibnamefont {Glenn}},\ and\
  \bibinfo {author} {\bibfnamefont {D.}~\bibnamefont {DeMille}},\ }\href
  {https://doi.org/10.1103/PhysRevLett.103.223001} {\bibfield  {journal}
  {\bibinfo  {journal} {Phys. Rev. Lett.}\ }\textbf {\bibinfo {volume} {103}},\
  \bibinfo {pages} {223001} (\bibinfo {year} {2009})}\BibitemShut {NoStop}%
\bibitem [{\citenamefont {Shuman}\ \emph {et~al.}(2010)\citenamefont {Shuman},
  \citenamefont {Barry},\ and\ \citenamefont
  {DeMille}}]{shuman2010laserlasercoolingmolecules}%
  \BibitemOpen
  \bibfield  {author} {\bibinfo {author} {\bibfnamefont {E.~S.}\ \bibnamefont
  {Shuman}}, \bibinfo {author} {\bibfnamefont {J.~F.}\ \bibnamefont {Barry}},\
  and\ \bibinfo {author} {\bibfnamefont {D.}~\bibnamefont {DeMille}},\ }\href
  {https://doi.org/10.1038/nature09443} {\bibfield  {journal} {\bibinfo
  {journal} {Nature}\ }\textbf {\bibinfo {volume} {467}},\ \bibinfo {pages}
  {820} (\bibinfo {year} {2010})}\BibitemShut {NoStop}%
\bibitem [{\citenamefont {Chan}\ \emph {et~al.}(2011)\citenamefont {Chan},
  \citenamefont {Alegre}, \citenamefont {Safavi-Naeini}, \citenamefont {Hill},
  \citenamefont {Krause}, \citenamefont {Gr{\"o}blacher}, \citenamefont
  {Aspelmeyer},\ and\ \citenamefont
  {Painter}}]{chan2011lasercooling_resonator}%
  \BibitemOpen
  \bibfield  {author} {\bibinfo {author} {\bibfnamefont {J.}~\bibnamefont
  {Chan}}, \bibinfo {author} {\bibfnamefont {T.}~\bibnamefont {Alegre}},
  \bibinfo {author} {\bibfnamefont {A.~H.}\ \bibnamefont {Safavi-Naeini}},
  \bibinfo {author} {\bibfnamefont {J.~T.}\ \bibnamefont {Hill}}, \bibinfo
  {author} {\bibfnamefont {A.}~\bibnamefont {Krause}}, \bibinfo {author}
  {\bibfnamefont {S.}~\bibnamefont {Gr{\"o}blacher}}, \bibinfo {author}
  {\bibfnamefont {M.}~\bibnamefont {Aspelmeyer}},\ and\ \bibinfo {author}
  {\bibfnamefont {O.}~\bibnamefont {Painter}},\ }\href
  {https://doi.org/10.1038/nature10461} {\bibfield  {journal} {\bibinfo
  {journal} {Nature}\ }\textbf {\bibinfo {volume} {478}},\ \bibinfo {pages}
  {89} (\bibinfo {year} {2011})}\BibitemShut {NoStop}%
\bibitem [{\citenamefont {Teufel}\ \emph {et~al.}(2011)\citenamefont {Teufel},
  \citenamefont {Donner}, \citenamefont {Li}, \citenamefont {Harlow},
  \citenamefont {Allman}, \citenamefont {Cicak}, \citenamefont {Sirois},
  \citenamefont {Whittaker}, \citenamefont {Lehnert},\ and\ \citenamefont
  {Simmonds}}]{teufel2011lasercooling_resonator}%
  \BibitemOpen
  \bibfield  {author} {\bibinfo {author} {\bibfnamefont {J.~D.}\ \bibnamefont
  {Teufel}}, \bibinfo {author} {\bibfnamefont {T.}~\bibnamefont {Donner}},
  \bibinfo {author} {\bibfnamefont {D.}~\bibnamefont {Li}}, \bibinfo {author}
  {\bibfnamefont {J.~W.}\ \bibnamefont {Harlow}}, \bibinfo {author}
  {\bibfnamefont {M.}~\bibnamefont {Allman}}, \bibinfo {author} {\bibfnamefont
  {K.}~\bibnamefont {Cicak}}, \bibinfo {author} {\bibfnamefont {A.~J.}\
  \bibnamefont {Sirois}}, \bibinfo {author} {\bibfnamefont {J.~D.}\
  \bibnamefont {Whittaker}}, \bibinfo {author} {\bibfnamefont {K.~W.}\
  \bibnamefont {Lehnert}},\ and\ \bibinfo {author} {\bibfnamefont {R.~W.}\
  \bibnamefont {Simmonds}},\ }\href {https://doi.org/10.1038/nature10261}
  {\bibfield  {journal} {\bibinfo  {journal} {Nature}\ }\textbf {\bibinfo
  {volume} {475}},\ \bibinfo {pages} {359} (\bibinfo {year}
  {2011})}\BibitemShut {NoStop}%
\bibitem [{\citenamefont {Caldwell}\ and\ \citenamefont
  {Tarbutt}(2020)}]{caldwell2020}%
  \BibitemOpen
  \bibfield  {author} {\bibinfo {author} {\bibfnamefont {L.}~\bibnamefont
  {Caldwell}}\ and\ \bibinfo {author} {\bibfnamefont {M.~R.}\ \bibnamefont
  {Tarbutt}},\ }\href {https://doi.org/10.1103/PhysRevResearch.2.013251}
  {\bibfield  {journal} {\bibinfo  {journal} {Phys. Rev. Res.}\ }\textbf
  {\bibinfo {volume} {2}},\ \bibinfo {pages} {013251} (\bibinfo {year}
  {2020})}\BibitemShut {NoStop}%
\bibitem [{\citenamefont {Vanner}\ \emph {et~al.}(2013)\citenamefont {Vanner},
  \citenamefont {Hofer}, \citenamefont {Cole},\ and\ \citenamefont
  {Aspelmeyer}}]{vanner2013}%
  \BibitemOpen
  \bibfield  {author} {\bibinfo {author} {\bibfnamefont {M.~R.}\ \bibnamefont
  {Vanner}}, \bibinfo {author} {\bibfnamefont {J.}~\bibnamefont {Hofer}},
  \bibinfo {author} {\bibfnamefont {G.~D.}\ \bibnamefont {Cole}},\ and\
  \bibinfo {author} {\bibfnamefont {M.}~\bibnamefont {Aspelmeyer}},\ }\href
  {https://doi.org/10.1038/ncomms3295} {\bibfield  {journal} {\bibinfo
  {journal} {Nature Comms.}\ }\textbf {\bibinfo {volume} {4}},\ \bibinfo
  {pages} {2295} (\bibinfo {year} {2013})}\BibitemShut {NoStop}%
\bibitem [{\citenamefont {Mooser}\ \emph {et~al.}(2013)\citenamefont {Mooser},
  \citenamefont {Bräuninger}, \citenamefont {Franke}, \citenamefont {Kracke},
  \citenamefont {Leiteritz}, \citenamefont {Rodegheri}, \citenamefont
  {Nagahama}, \citenamefont {Schneider}, \citenamefont {Smorra}, \citenamefont
  {Blaum}, \citenamefont {Matsuda}, \citenamefont {Quint}, \citenamefont
  {Walz}, \citenamefont {Yamazaki},\ and\ \citenamefont {Ulmer}}]{mooser2013}%
  \BibitemOpen
  \bibfield  {author} {\bibinfo {author} {\bibfnamefont {A.}~\bibnamefont
  {Mooser}}, \bibinfo {author} {\bibfnamefont {S.}~\bibnamefont {Bräuninger}},
  \bibinfo {author} {\bibfnamefont {K.}~\bibnamefont {Franke}}, \bibinfo
  {author} {\bibfnamefont {H.}~\bibnamefont {Kracke}}, \bibinfo {author}
  {\bibfnamefont {C.}~\bibnamefont {Leiteritz}}, \bibinfo {author}
  {\bibfnamefont {C.}~\bibnamefont {Rodegheri}}, \bibinfo {author}
  {\bibfnamefont {H.}~\bibnamefont {Nagahama}}, \bibinfo {author}
  {\bibfnamefont {G.}~\bibnamefont {Schneider}}, \bibinfo {author}
  {\bibfnamefont {C.}~\bibnamefont {Smorra}}, \bibinfo {author} {\bibfnamefont
  {K.}~\bibnamefont {Blaum}}, \bibinfo {author} {\bibfnamefont
  {Y.}~\bibnamefont {Matsuda}}, \bibinfo {author} {\bibfnamefont
  {W.}~\bibnamefont {Quint}}, \bibinfo {author} {\bibfnamefont
  {J.}~\bibnamefont {Walz}}, \bibinfo {author} {\bibfnamefont {Y.}~\bibnamefont
  {Yamazaki}},\ and\ \bibinfo {author} {\bibfnamefont {S.}~\bibnamefont
  {Ulmer}},\ }\href
  {https://doi.org/https://doi.org/10.1016/j.physletb.2013.05.012} {\bibfield
  {journal} {\bibinfo  {journal} {Physics Letters B}\ }\textbf {\bibinfo
  {volume} {723}},\ \bibinfo {pages} {78} (\bibinfo {year} {2013})}\BibitemShut
  {NoStop}%
\bibitem [{\citenamefont {Puebla}\ \emph {et~al.}(2020)\citenamefont {Puebla},
  \citenamefont {Abah},\ and\ \citenamefont
  {Paternostro}}]{puebla2020measurementcooling}%
  \BibitemOpen
  \bibfield  {author} {\bibinfo {author} {\bibfnamefont {R.}~\bibnamefont
  {Puebla}}, \bibinfo {author} {\bibfnamefont {O.}~\bibnamefont {Abah}},\ and\
  \bibinfo {author} {\bibfnamefont {M.}~\bibnamefont {Paternostro}},\ }\href
  {https://doi.org/10.1103/PhysRevB.101.245410} {\bibfield  {journal} {\bibinfo
   {journal} {Phys. Rev. B}\ }\textbf {\bibinfo {volume} {101}},\ \bibinfo
  {pages} {245410} (\bibinfo {year} {2020})}\BibitemShut {NoStop}%
\bibitem [{\citenamefont {Montenegro}\ \emph {et~al.}(2018)\citenamefont
  {Montenegro}, \citenamefont {Coto}, \citenamefont {Eremeev},\ and\
  \citenamefont {Orszag}}]{Montenegro2018measurementcooling}%
  \BibitemOpen
  \bibfield  {author} {\bibinfo {author} {\bibfnamefont {V.}~\bibnamefont
  {Montenegro}}, \bibinfo {author} {\bibfnamefont {R.}~\bibnamefont {Coto}},
  \bibinfo {author} {\bibfnamefont {V.}~\bibnamefont {Eremeev}},\ and\ \bibinfo
  {author} {\bibfnamefont {M.}~\bibnamefont {Orszag}},\ }\href
  {https://doi.org/10.1103/PhysRevA.98.053837} {\bibfield  {journal} {\bibinfo
  {journal} {Phys. Rev. A}\ }\textbf {\bibinfo {volume} {98}},\ \bibinfo
  {pages} {053837} (\bibinfo {year} {2018})}\BibitemShut {NoStop}%
\bibitem [{\citenamefont {Bergenfeldt}\ and\ \citenamefont
  {M\o{}lmer}(2009)}]{Bergenfeldt2009measurementcooling}%
  \BibitemOpen
  \bibfield  {author} {\bibinfo {author} {\bibfnamefont {C.}~\bibnamefont
  {Bergenfeldt}}\ and\ \bibinfo {author} {\bibfnamefont {K.}~\bibnamefont
  {M\o{}lmer}},\ }\href {https://doi.org/10.1103/PhysRevA.80.043838} {\bibfield
   {journal} {\bibinfo  {journal} {Phys. Rev. A}\ }\textbf {\bibinfo {volume}
  {80}},\ \bibinfo {pages} {043838} (\bibinfo {year} {2009})}\BibitemShut
  {NoStop}%
\bibitem [{\citenamefont {Vogelius}\ \emph {et~al.}(2006)\citenamefont
  {Vogelius}, \citenamefont {Madsen},\ and\ \citenamefont
  {Drewsen}}]{vogelius2006}%
  \BibitemOpen
  \bibfield  {author} {\bibinfo {author} {\bibfnamefont {I.~S.}\ \bibnamefont
  {Vogelius}}, \bibinfo {author} {\bibfnamefont {L.~B.}\ \bibnamefont
  {Madsen}},\ and\ \bibinfo {author} {\bibfnamefont {M.}~\bibnamefont
  {Drewsen}},\ }\href {https://doi.org/10.1088/0953-4075/39/19/S31} {\bibfield
  {journal} {\bibinfo  {journal} {Journal of Physics B: Atomic, Molecular and
  Optical Physics}\ }\textbf {\bibinfo {volume} {39}},\ \bibinfo {pages}
  {S1259} (\bibinfo {year} {2006})}\BibitemShut {NoStop}%
\bibitem [{\citenamefont {Chou}\ \emph {et~al.}(2017)\citenamefont {Chou},
  \citenamefont {Kurz}, \citenamefont {Hume}, \citenamefont {Plessow},
  \citenamefont {Leibrandt},\ and\ \citenamefont {Leibfried}}]{chou2017}%
  \BibitemOpen
  \bibfield  {author} {\bibinfo {author} {\bibfnamefont {C.-w.}\ \bibnamefont
  {Chou}}, \bibinfo {author} {\bibfnamefont {C.}~\bibnamefont {Kurz}}, \bibinfo
  {author} {\bibfnamefont {D.~B.}\ \bibnamefont {Hume}}, \bibinfo {author}
  {\bibfnamefont {P.~N.}\ \bibnamefont {Plessow}}, \bibinfo {author}
  {\bibfnamefont {D.~R.}\ \bibnamefont {Leibrandt}},\ and\ \bibinfo {author}
  {\bibfnamefont {D.}~\bibnamefont {Leibfried}},\ }\href
  {https://doi.org/10.1038/nature22338} {\bibfield  {journal} {\bibinfo
  {journal} {Nature}\ }\textbf {\bibinfo {volume} {545}},\ \bibinfo {pages}
  {203} (\bibinfo {year} {2017})}\BibitemShut {NoStop}%
\bibitem [{\citenamefont {Allen}\ and\ \citenamefont
  {Eberly}(1987)}]{allen1987optical}%
  \BibitemOpen
  \bibfield  {author} {\bibinfo {author} {\bibfnamefont {L.}~\bibnamefont
  {Allen}}\ and\ \bibinfo {author} {\bibfnamefont {J.~H.}\ \bibnamefont
  {Eberly}},\ }\href@noop {} {\emph {\bibinfo {title} {Optical resonance and
  two-level atoms}}},\ Vol.~\bibinfo {volume} {28}\ (\bibinfo  {publisher}
  {Courier Corporation},\ \bibinfo {year} {1987})\BibitemShut {NoStop}%
\bibitem [{\citenamefont {Corfield}\ \emph {et~al.}(2021)\citenamefont
  {Corfield}, \citenamefont {Lishman}, \citenamefont {Lee}, \citenamefont
  {Toba}, \citenamefont {Porter}, \citenamefont {Heinrich}, \citenamefont
  {Webster}, \citenamefont {Mintert},\ and\ \citenamefont
  {Thompson}}]{corfield2021}%
  \BibitemOpen
  \bibfield  {author} {\bibinfo {author} {\bibfnamefont {O.}~\bibnamefont
  {Corfield}}, \bibinfo {author} {\bibfnamefont {J.}~\bibnamefont {Lishman}},
  \bibinfo {author} {\bibfnamefont {C.}~\bibnamefont {Lee}}, \bibinfo {author}
  {\bibfnamefont {J.~M.}\ \bibnamefont {Toba}}, \bibinfo {author}
  {\bibfnamefont {G.}~\bibnamefont {Porter}}, \bibinfo {author} {\bibfnamefont
  {J.~M.}\ \bibnamefont {Heinrich}}, \bibinfo {author} {\bibfnamefont {S.~C.}\
  \bibnamefont {Webster}}, \bibinfo {author} {\bibfnamefont {F.}~\bibnamefont
  {Mintert}},\ and\ \bibinfo {author} {\bibfnamefont {R.~C.}\ \bibnamefont
  {Thompson}},\ }\href {https://doi.org/10.1103/PRXQuantum.2.040359} {\bibfield
   {journal} {\bibinfo  {journal} {PRX Quantum}\ }\textbf {\bibinfo {volume}
  {2}},\ \bibinfo {pages} {040359} (\bibinfo {year} {2021})}\BibitemShut
  {NoStop}%
\bibitem [{\citenamefont {Gulde}(2003)}]{Gulde2003}%
  \BibitemOpen
  \bibfield  {author} {\bibinfo {author} {\bibfnamefont {S.}~\bibnamefont
  {Gulde}},\ }\emph {\bibinfo {title} {Experimental Realisation of Quantum
  Gates and the Deutsch-Jozsa Algorithm with Trapped $^{40}$Ca$^+$ Ions}},\
  \href {https://quantumoptics.at/en/publications/ph-d-theses.html} {Ph.D.
  thesis},\ \bibinfo  {school} {Universit\"at Innsbruck} (\bibinfo {year}
  {2003})\BibitemShut {NoStop}%
\bibitem [{\citenamefont {Leibfried}\ \emph {et~al.}(2003)\citenamefont
  {Leibfried}, \citenamefont {Blatt}, \citenamefont {Monroe},\ and\
  \citenamefont {Wineland}}]{Leibfried2003}%
  \BibitemOpen
  \bibfield  {author} {\bibinfo {author} {\bibfnamefont {D.}~\bibnamefont
  {Leibfried}}, \bibinfo {author} {\bibfnamefont {R.}~\bibnamefont {Blatt}},
  \bibinfo {author} {\bibfnamefont {C.}~\bibnamefont {Monroe}},\ and\ \bibinfo
  {author} {\bibfnamefont {D.~J.}\ \bibnamefont {Wineland}},\ }\href
  {https://doi.org/10.1103/RevModPhys.75.281} {\bibfield  {journal} {\bibinfo
  {journal} {Review of Modern Physics}\ }\textbf {\bibinfo {volume} {75}},\
  \bibinfo {pages} {281} (\bibinfo {year} {2003})}\BibitemShut {NoStop}%
\bibitem [{\citenamefont {Wineland}\ \emph {et~al.}(1998)\citenamefont
  {Wineland}, \citenamefont {Monroe}, \citenamefont {Itano}, \citenamefont
  {Leibfried}, \citenamefont {King},\ and\ \citenamefont
  {Meekhav}}]{wineland1988}%
  \BibitemOpen
  \bibfield  {author} {\bibinfo {author} {\bibfnamefont {D.~J.}\ \bibnamefont
  {Wineland}}, \bibinfo {author} {\bibfnamefont {C.}~\bibnamefont {Monroe}},
  \bibinfo {author} {\bibfnamefont {W.~M.}\ \bibnamefont {Itano}}, \bibinfo
  {author} {\bibfnamefont {D.}~\bibnamefont {Leibfried}}, \bibinfo {author}
  {\bibfnamefont {B.~E.}\ \bibnamefont {King}},\ and\ \bibinfo {author}
  {\bibfnamefont {D.~M.}\ \bibnamefont {Meekhav}},\ }\href
  {https://doi.org/10.6028/jres.103.019} {\bibfield  {journal} {\bibinfo
  {journal} {J. Res. Natl. Inst. Stand. Technol.}\ }\textbf {\bibinfo {volume}
  {103}},\ \bibinfo {pages} {259} (\bibinfo {year} {1998})}\BibitemShut
  {NoStop}%
\bibitem [{\citenamefont {Watanabe}\ \emph {et~al.}(2011)\citenamefont
  {Watanabe}, \citenamefont {Nomura}, \citenamefont {Toyoda},\ and\
  \citenamefont {Urabe}}]{Watanabe2011RAPsideband}%
  \BibitemOpen
  \bibfield  {author} {\bibinfo {author} {\bibfnamefont {T.}~\bibnamefont
  {Watanabe}}, \bibinfo {author} {\bibfnamefont {S.}~\bibnamefont {Nomura}},
  \bibinfo {author} {\bibfnamefont {K.}~\bibnamefont {Toyoda}},\ and\ \bibinfo
  {author} {\bibfnamefont {S.}~\bibnamefont {Urabe}},\ }\href
  {https://doi.org/10.1103/PhysRevA.84.033412} {\bibfield  {journal} {\bibinfo
  {journal} {Phys. Rev. A}\ }\textbf {\bibinfo {volume} {84}},\ \bibinfo
  {pages} {033412} (\bibinfo {year} {2011})}\BibitemShut {NoStop}%
\bibitem [{\citenamefont {Wunderlich}\ \emph {et~al.}(2007)\citenamefont
  {Wunderlich}, \citenamefont {Hannemann}, \citenamefont {Körber},
  \citenamefont {Häffner}, \citenamefont {Roos}, \citenamefont {Hänsel},
  \citenamefont {Blatt},\ and\ \citenamefont
  {Schmidt-Kaler}}]{WunderlichChr2007RAPiontrap}%
  \BibitemOpen
  \bibfield  {author} {\bibinfo {author} {\bibfnamefont {C.}~\bibnamefont
  {Wunderlich}}, \bibinfo {author} {\bibfnamefont {T.}~\bibnamefont
  {Hannemann}}, \bibinfo {author} {\bibfnamefont {T.}~\bibnamefont {Körber}},
  \bibinfo {author} {\bibfnamefont {H.}~\bibnamefont {Häffner}}, \bibinfo
  {author} {\bibfnamefont {C.}~\bibnamefont {Roos}}, \bibinfo {author}
  {\bibfnamefont {W.}~\bibnamefont {Hänsel}}, \bibinfo {author} {\bibfnamefont
  {R.}~\bibnamefont {Blatt}},\ and\ \bibinfo {author} {\bibfnamefont
  {F.}~\bibnamefont {Schmidt-Kaler}},\ }\href
  {https://doi.org/10.1080/09500340600741082} {\bibfield  {journal} {\bibinfo
  {journal} {Journal of modern optics}\ }\textbf {\bibinfo {volume} {54}},\
  \bibinfo {pages} {1541} (\bibinfo {year} {2007})}\BibitemShut {NoStop}%
\bibitem [{\citenamefont {Toyoda}\ \emph {et~al.}(2011)\citenamefont {Toyoda},
  \citenamefont {Watanabe}, \citenamefont {Kimura}, \citenamefont {Nomura},
  \citenamefont {Haze},\ and\ \citenamefont {Urabe}}]{Toyoda2011RAP_on_SB}%
  \BibitemOpen
  \bibfield  {author} {\bibinfo {author} {\bibfnamefont {K.}~\bibnamefont
  {Toyoda}}, \bibinfo {author} {\bibfnamefont {T.}~\bibnamefont {Watanabe}},
  \bibinfo {author} {\bibfnamefont {T.}~\bibnamefont {Kimura}}, \bibinfo
  {author} {\bibfnamefont {S.}~\bibnamefont {Nomura}}, \bibinfo {author}
  {\bibfnamefont {S.}~\bibnamefont {Haze}},\ and\ \bibinfo {author}
  {\bibfnamefont {S.}~\bibnamefont {Urabe}},\ }\href
  {https://doi.org/10.1103/PhysRevA.83.022315} {\bibfield  {journal} {\bibinfo
  {journal} {Phys. Rev. A}\ }\textbf {\bibinfo {volume} {83}},\ \bibinfo
  {pages} {022315} (\bibinfo {year} {2011})}\BibitemShut {NoStop}%
\bibitem [{\citenamefont {Lu}\ \emph {et~al.}(2007)\citenamefont {Lu},
  \citenamefont {Miao},\ and\ \citenamefont
  {Metcalf}}]{Lu2007non_adiabatic_transition}%
  \BibitemOpen
  \bibfield  {author} {\bibinfo {author} {\bibfnamefont {T.}~\bibnamefont
  {Lu}}, \bibinfo {author} {\bibfnamefont {X.}~\bibnamefont {Miao}},\ and\
  \bibinfo {author} {\bibfnamefont {H.}~\bibnamefont {Metcalf}},\ }\href
  {https://doi.org/10.1103/PhysRevA.75.063422} {\bibfield  {journal} {\bibinfo
  {journal} {Phys. Rev. A}\ }\textbf {\bibinfo {volume} {75}},\ \bibinfo
  {pages} {063422} (\bibinfo {year} {2007})}\BibitemShut {NoStop}%
\bibitem [{\citenamefont {Wineland}\ and\ \citenamefont
  {Itano}(1979)}]{Wineland1979state_dependent_rabi}%
  \BibitemOpen
  \bibfield  {author} {\bibinfo {author} {\bibfnamefont {D.~J.}\ \bibnamefont
  {Wineland}}\ and\ \bibinfo {author} {\bibfnamefont {W.~M.}\ \bibnamefont
  {Itano}},\ }\href {https://doi.org/10.1103/PhysRevA.20.1521} {\bibfield
  {journal} {\bibinfo  {journal} {Phys. Rev. A}\ }\textbf {\bibinfo {volume}
  {20}},\ \bibinfo {pages} {1521} (\bibinfo {year} {1979})}\BibitemShut
  {NoStop}%
\bibitem [{\citenamefont {Arcizet}\ \emph {et~al.}(2011)\citenamefont
  {Arcizet}, \citenamefont {Jacques}, \citenamefont {Siria}, \citenamefont
  {Poncharal}, \citenamefont {Vincent},\ and\ \citenamefont
  {Seidelin}}]{arcizet2011}%
  \BibitemOpen
  \bibfield  {author} {\bibinfo {author} {\bibfnamefont {O.}~\bibnamefont
  {Arcizet}}, \bibinfo {author} {\bibfnamefont {V.}~\bibnamefont {Jacques}},
  \bibinfo {author} {\bibfnamefont {A.}~\bibnamefont {Siria}}, \bibinfo
  {author} {\bibfnamefont {P.}~\bibnamefont {Poncharal}}, \bibinfo {author}
  {\bibfnamefont {P.}~\bibnamefont {Vincent}},\ and\ \bibinfo {author}
  {\bibfnamefont {S.}~\bibnamefont {Seidelin}},\ }\href
  {https://doi.org/10.1038/nphys2070} {\bibfield  {journal} {\bibinfo
  {journal} {Nature Physics}\ }\textbf {\bibinfo {volume} {7}},\ \bibinfo
  {pages} {879} (\bibinfo {year} {2011})}\BibitemShut {NoStop}%
\end{thebibliography}%

\clearpage

\end{document}